\documentclass[preprint,aps]{revtex4}
 \usepackage{rotating}

\usepackage{graphicx}
 
\begin{document}
 \newcommand{\Qed}{\rule{2.5mm}{3mm}}
 \newcommand{\balpha}{\mbox{\boldmath {$\alpha$}}}
 \draft 
%\documentstyle[preprint,aps]{revtex}
%\begin{document}
%
%
\title{Quantum gates and quantum algorithms with Clifford algebra technique} 
\author{ M. Gregori\v c,  
N.S. Manko\v c Bor\v stnik\\
%\address{ 
Department of Physics, University of
Ljubljana, Jadranska 19, 1000 Ljubljana,\\
}
\date{\today}
\begin{abstract} 
We use the Clifford algebra technique~\cite{hn02,hn03}, that is  
nilpotents and projectors which are binomials  of the Clifford algebra 
objects $\gamma^a$ with the property $\{\gamma^a,\gamma^b\}_+ = 2 \eta^{ab}$,  
for representing 
 quantum gates and quantum algorithms needed in  quantum computers 
 in an elegant way. 
We identify $n$-qubits with  spinor 
representations of the group $SO(1,3)$ for a system of $n$ spinors. Representations  
are expressed 
in terms of products of projectors and nilpotents. An algorithm for 
extracting a particular information out of  a general superposition of $2^n$ qubit states 
is presented. It 
reproduces for a particular choice of the initial state  the Grover's 
algorithm~\cite{grover}. 
\end{abstract}

\maketitle

\section{Introduction}
\label{introduction}

It is easy to prove (and it is also well known) 
that any type of a 
quantum gate, operating on one qubit and represented by an unitary operator, can be expressed 
as a product of the two types of quantum gates---the phase gate and the Hadamard's gate---while 
the C-NOT gate, operating on two quantum bits, enables to make a quantum computer realizable, 
since all the needed operations can be expressed in terms of these three types of gates. 
In the references~\cite{a,b} the use of the geometrical 
algebra  to demonstrate these gates and their functioning  is presented.

In this paper we use the technique from the references~\cite{hn02,hn03}, which represents spinor 
representations of the group $SO(1,3)$ in terms of projectors and nilpotents, which 
are binomials of the Clifford algebra objects $\gamma^a$ and we   
 identify the spinor representation of two one spinor states with the 
two quantum bits $|0 \rangle$ and $|1 \rangle$ and accordingly 
$n$ spinors' representation  of $SO(1,3)$  with the $n$-qubits.  
The three types of the gates can then be expressed in 
terms of projectors and nilpotents in a transparent and elegant way. 

We present also the %general 
algorithm 
for extracting a particular information out of any superposition of a $n$-qubit state. 
For a particular choice of the initial  $n$-qubit state this general algorithm 
reproduces the
Grover's algorithm~\cite{grover}.

\section{The technique for  spinor representations} 
\label{spinorstates}

We define in this section ($2^{4/2-1}$) 
basic states  for the spinor (fundamental) 
representation of the group $SO(1,3)$ by expressing  the basic states as polynomials 
of the Clifford algebra objects $\gamma^a$---nilpotents and 
projectors~\cite{hn02,hn03}---chosen to be eigenvectors of the Cartan subalgebra set 
(with $2$ commuting operators) of the $6$ infinitesimal operators of the group $SO(1,3)$. 
We identify  one qubit with one of the two spinor basic states, distinguishing 
between the chiral representation and the representation with a well defined parity. 
We make at the end use of  states with well defined parity, %since they seem to 
%be more appropriate for the realizable types of quantum computers, 
although both representations for spinors are equivalent and  the 
proposed quantum gates work for  both representations.  
We identify $n$-qubits with states which are superposition of $2^n$ 
products of $n$ one spinor states. 
We also present some relations, useful when defining the quantum gates.

The six infinitesimal generators $S^{ab}$ of the group $SO(1,3)$ 
($S^{01}$, $S^{02}$, $S^{03}$, $S^{23}$, $S^{31}$, $S^{12}$) fulfill 
the Lorentz algebra $\{S^{ab},S^{cd}\}_- = i( \eta^{ad} S^{bc} + 
\eta^{bc} S^{ad} - \eta^{ac} S^{bd} - \eta^{bd} S^{ac})$.  For spinors can the generators 
$ S^{ab}$ be written in terms of the operators $\gamma^a$ fulfilling the Clifford algebra
\begin{eqnarray}
\{\gamma^a,\gamma^b\}_+ = 2 \eta^{ab}, \quad {\rm diag}(\eta)= (1,-1,-1,-1),\nonumber\\ 
S^{ab} = \frac{i}{2} \gamma^a \gamma^b, \;\; {\rm for} \, a\ne b \;{\rm and}\; 0\; {\rm otherwise}.
\label{clifford}
\end{eqnarray}
They define the spinor (fundamental) representation of the group $SO(1,3)$.  
Choosing for the Cartan subalgebra set of commuting operators $S^{03}$ and $S^{12}$, 
the spinor states 
\begin{eqnarray}
|0\rangle_L &=& \stackrel{03}{[-i]}\stackrel{12}{(+)}|\varphi_0\rangle,\quad 
|1\rangle_L = \stackrel{03}{(+i)}\stackrel{12}{[-]}|\varphi_0\rangle,\nonumber\\
|0\rangle_R &=& \stackrel{03}{(+i)}\stackrel{12}{(+)}|\varphi_0\rangle,\quad  
|1\rangle_R = \stackrel{03}{[-i]}\stackrel{12}{[-]}|\varphi_0\rangle,
\label{spinorstatesh}
\end{eqnarray}
where $|\varphi_0\rangle$ is a vacuum state\cite{hn02,hn03} 
(which from now on we shall skip it since its only required property is that 
when the Clifford algebra objects from Eq.(\ref{spinorstatesh}) are applied on the vacuum state 
the left hand side do not give zero)
and with nilpotents ($\stackrel{ab}{(k)} \stackrel{ab}{(k)}=0$) and  projectors 
($\stackrel{ab}{[k]} \stackrel{ab}{[k]}=
\stackrel{ab}{[k]}$) defined as
\begin{eqnarray}
 \stackrel{03}{(\pm i)} &:=& \frac{1}{2} (\gamma^0 \mp \gamma^3), \quad 
 \stackrel{12}{(\pm)} := \frac{1}{2} (\gamma^1 \pm i \gamma^2),\nonumber\\ 
\stackrel{03}{[\pm i]} &:=& \frac{1}{2} (1\pm \gamma^0\gamma^3), \quad 
 \stackrel{12}{[\pm]} := \frac{1}{2} (1 \pm i \gamma^1 \gamma^2),
\label{spinorstatesh1}
\end{eqnarray}
are all eigenstates of the Cartan subalgbra  set $S^{03}$ and $S^{12}$, since  
 $S^{03} \stackrel{03}{(\pm i)} = \pm \frac{i}{2} \stackrel{03}{(\pm i)}
 $, $S^{03} \stackrel{03}{[\pm i]} = \pm \frac{i}{2} \stackrel{03}{[\pm i]}
 $ and similarly  $S^{12} \stackrel{12}{(\pm )} = \pm \frac{1}{2} \stackrel{12}{(\pm )}
 $, $S^{12} \stackrel{12}{[\pm ]} = \pm \frac{1}{2} \stackrel{12}{[\pm ]}
 $, what can very easily be checked just by applying $S^{03}$ and $S^{12}$ on a  
 particular nilpotent ($\stackrel{03}{(\pm i)},  \stackrel{12}{(\pm)}$) or projector 
 $\stackrel{03}{[\pm i]},\stackrel{12}{[\pm ]}$ and by 
 using Eq.(\ref{clifford}). The states $|0\rangle_L$ and $|1\rangle_L$  have handedness 
 $\Gamma= -4i S^{03}S^{12} $ equal to $-1$, while the states $|0\rangle_R$ and $|1\rangle_R$
 have handedness equal to $1$, which again can easily be proved just by inspection. 
 We normalize the states as follows \cite{hn02}
 \begin{eqnarray}
 {}_{\beta}\langle i|j\rangle_{\alpha} = \delta_{ij} \delta_{\alpha \beta}, 
  \label{spinorstatesnor}
 \end{eqnarray}
 where $i,j$ denote $0$ or $1$ and $\alpha,\beta$ left and right handedness. 
 
 When describing a spinor in its center of mass motion, the representation 
 with a well defined parity is more convenient 
 \begin{eqnarray}
 |0\rangle &=& \frac{1}{\sqrt{2}}(\stackrel{03}{[-i]}\stackrel{12}{(+)} \pm 
 \stackrel{03}{(+i)}\stackrel{12}{(+)}),\nonumber\\
 |1\rangle &=& \frac{1}{\sqrt{2}}(\stackrel{03}{(+i)}\stackrel{12}{[-]} \pm
 \stackrel{03}{[-i]}\stackrel{12}{[-]}).
 \label{spinorstatesp}
 \end{eqnarray}
Usually spinors with a positive parity are identified with particles, while 
spinors with a negative parity are identified with antiparticles. For our purpose 
parity property does not matter. We can use any of these two types of states. 
Nilpotents and projectors fulfil the following relations \cite{hn02,hn03} 
(which can be checked just by using the definition of the nilpotents 
and projectors (Eq.\ref{spinorstatesh1}) and by taking into account 
the Clifford property of $\gamma^a$'s (Eq.\ref{clifford}))
\begin{eqnarray}
\stackrel{ab}{(k)}\stackrel{ab}{(k)}& =& 0, \quad \quad \stackrel{ab}{(k)}\stackrel{ab}{(-k)}
= \eta^{aa}  \stackrel{ab}{[k]}, \quad %\stackrel{ab}{(-k)}\stackrel{ab}{(k)}=
\stackrel{ab}{[k]}\stackrel{ab}{[k]} =  \stackrel{ab}{[k]}, \quad \quad
\stackrel{ab}{[k]}\stackrel{ab}{[-k]}= 0,  \nonumber\\
\stackrel{ab}{(k)}\stackrel{ab}{[k]}& =& 0,\quad \quad \quad \stackrel{ab}{[k]}\stackrel{ab}{(k)}
=  \stackrel{ab}{(k)}, \quad \quad 
\stackrel{ab}{(k)}\stackrel{ab}{[-k]} =  \stackrel{ab}{(k)},
\quad \quad \stackrel{ab}{[k]}\stackrel{ab}{(-k)} =0.   
\label{raiselower}
\end{eqnarray}
We then find that the operators 
\begin{eqnarray}
\tau_{L}^{\mp} &:=& - \stackrel{03}{(\pm i)}\stackrel{12}{(\mp)},\quad 
\tau_{R}^{\mp} :=  \stackrel{03}{(\mp i)}\stackrel{12}{(\mp)}  
\label{taulr}
\end{eqnarray}
transform a state of a particular handedness (left or right)  
 into a state of the same handedness or annihilate it, 
 while they annihilate the states of an opposite handedness
\begin{eqnarray}
\tau_{L}^{-} |0\rangle_L &=& |1\rangle_L, \quad \tau_{L}^{+} |1\rangle_L = |0\rangle_L,  
\nonumber\\
\tau_{R}^{-} |0\rangle_R &=& |1\rangle_R, \quad \tau_{R}^{+} |1\rangle_R = |0\rangle_R.
\label{taulrstates}
\end{eqnarray}
All the other applications $\tau_{L}^{\mp} $ and $\tau_{R}^{\mp}$  give zero. 

We also find that the operators 
\begin{eqnarray}
\tau^{\mp}: &=& \tau_{L}^{\mp} + \tau_{R}^{\mp}= - \stackrel{03}{(\pm i)}\stackrel{12}{(\mp)}
 +  \stackrel{03}{(\mp i)}\stackrel{12}{(\mp)}
\label{tau}
\end{eqnarray}
transform  a state of a well defined parity (Eq.\ref{spinorstatesp}) 
into a state of the same parity  or annihilate it 
 \begin{eqnarray}
 \tau^{-}|0\rangle &=& |1\rangle, \quad  \tau^{+}|1\rangle = |0\rangle,  
 \label{taustates}
 \end{eqnarray}
while the rest of applications give zero, accordingly $(\tau^{+}+\tau^{-})|0\rangle = |1\rangle, 
(\tau^{+}+\tau^{-})|1\rangle = |0\rangle$.

We present the following useful properties of $\tau^{\pm}$, valid  for $\tau_{L}^{\pm}$ and 
$\tau_{R}^{\pm}$ as well,  so that we shall from now on skip the index $L,R,$
 \begin{eqnarray}
 (\tau^{\pm})^2&=&0,\quad (\tau^{\pm})^\dagger = \tau^{\mp},\quad 
 \tau^{+}\tau^{-} = \stackrel{12}{[+]}, \quad  \tau^{-}\tau^{+}= \stackrel{12}{[-]},\quad 
 (\tau^{+}+\tau^{-})^2 =I,\nonumber\\
 \tau^{+} \stackrel{12}{[+]} &=&0, \quad \;\;\,  \tau^{-} \stackrel{12}{[-]} =0, \quad\;\;
 \stackrel{12}{[+]}\tau^{-}=0, \quad \;\, \stackrel{12}{[-]}\tau^{+} = 0, \nonumber\\
 \tau^{+} \stackrel{12}{[-]} &=&\tau^{+}, \quad   \tau^{-} \stackrel{12}{[+]} = \tau^{-}, \quad
 \stackrel{12}{[+]}\tau^{+}= \tau^{+}, \quad  \stackrel{12}{[-]}\tau^{-} = \tau^{-}.
 \label{taurel}
 \end{eqnarray}

A $n$-qubit state can be written in the chiral representation as
 \begin{eqnarray}
 |i_1 i_2 \cdots i_l \cdots i_n \rangle_{\alpha} &=& \prod_{l=1,n} |i_l \rangle_{\alpha},
 \quad \alpha = L,R,
  \label{nstateslr}
 \end{eqnarray}
while in the representation with  well defined parity we similarly have
 \begin{eqnarray}
 |i_1 i_2 \cdots i_l \cdots i_n \rangle &=& \prod_{l=1,n} |i_k \rangle.
  \label{nstates}
 \end{eqnarray}
$i_l$ stands for $|0 \rangle_l$ or $|1 \rangle_l$. 
All the raising and lowering operators $\tau^{\alpha \pm}_{l}$, $\alpha = L,R$ or 
$\tau^{ \pm}_{l}$ carry the index of  the corresponding qubit manifesting that 
they only apply on the particular $k$ state, while they do not ''see'' all the 
other states. Since they are made out of an even number of the Clifford odd nilpotents, they 
do not bring any sign when jumping over one-qubit states. 

Either in the chiral representation or in the representation with well defined parity 
basic states are the eigenstates of the operators $S^{12}$. According to 
Eq.(\ref{taustates}) any $n$-qubit state can be written as follows 
 \begin{eqnarray}
 |i_1 i_2 \cdots i_l \cdots i_n \rangle_{\alpha} &=& \prod_{l=1,n} 
 \,(\tau^{-}_{l})^{i_l}|0 \rangle_{l},
  \label{nstatesany}
 \end{eqnarray}
with $i_l$ equal to $0$ for the state with the eigenvalue of $S^{12}$ equal to $\frac{1}{2}$ or $1$ 
for the state with the eigenvalue of $S^{12}$ equal to $-\frac{1}{2}$.

\section{Quantum gates}
\label{quantumgates}

We define in this section three kinds of quantum gates: the phase gate and the 
Hadamard's gate, which apply on a particular qubit $l$ and the C-NOT gate, which applies 
on two qubits, say $l$ and $m$. All three gates are expressed in terms of projectors 
and an even number of nilpotents.

\vspace{3mm}

\noindent
i. The phase gate  ${\cal R}_{\Phi_l}$ is defined as
 \begin{eqnarray}
 {\cal R}_{\Phi_l }= \stackrel{12}{[+]}_l + e^{i \Phi_l} \stackrel{12}{[-]}_l.
  \label{phigate}
 \end{eqnarray}

{\it Statement:} The phase gate  ${\cal R}_{\Phi_l}$ if applying on  
$|0_l \rangle$ leaves it in the same state $|0_l \rangle$ without any change, while 
if applying on $|1_l \rangle$  multiplies this state with $e^{i\Phi}$. This is true 
for states with well defined parity $| i_l\rangle$ and also for the 
states with well defined handedness (%the chiral representation %
$|i_l \rangle_L$ 
and $|i_l \rangle_R$). 

{\it Proof:} To prove this statement one only has to apply the operator ${\cal R}_{\Phi_l}$
on $|i_l \rangle$, $|i_l \rangle_L$ and $|i_l \rangle_R$, with $i_l$ equal to $0$ or $1$, taking 
into account equations from Sect.~\ref{spinorstates}.

\vspace{3mm}

\noindent
ii. The Hadamard's gate ${\cal H}_l$ is defined as
 \begin{eqnarray}
 {\cal H}_l = \frac{1}{\sqrt{2}} [\stackrel{12}{[+]}_l - \stackrel{12}{[-]}_l - 
 \stackrel{03}{(+i)}_l  \stackrel{12}{(-)}_l + \stackrel{03}{(-i)}_l  \stackrel{12}{(-)}_l - 
 \stackrel{03}{(-i)}_l \stackrel{12}{(+)}_l + \stackrel{03}{(+i)}_l  \stackrel{12}{(+)}_l], 
  \label{hgate}
 \end{eqnarray}
or equivalently in terms of $\tau^{\pm}$ (Eq.(\ref{tau}))
 \begin{eqnarray}
 {\cal H}_l = \frac{1}{\sqrt{2}} [\stackrel{12}{[+]}_l - \stackrel{12}{[-]}_l + \tau^{-}_l 
 + \tau^{+}_l]. 
  \label{hgatetau}
 \end{eqnarray}

{\it Statement:} The Hadamard's gate  ${\cal H}_l$ if applying on  
$|0_l \rangle$ transforms it to ($\frac{1}{\sqrt{2}} (|0_l \rangle + |1_l \rangle)$), while 
if applying on $|1_l \rangle$  it transforms the state to 
($\frac{1}{\sqrt{2}} (|0_l \rangle - |1_l \rangle)$). This is true 
for states with well defined parity $| i_l\rangle$ and also for the 
states in the chiral representation $|i_l \rangle_L$ and $|i_l \rangle_R$. 

{\it Proof:} To prove this statement one only has to apply the operator ${\cal H}_l$
on $|i_l \rangle$, $|i_l \rangle_L$ and $|i_l \rangle_R$, with $i_l$ equal $0$ or $1$, 
taking into 
account  equations from Sect. \ref{spinorstates}.

\vspace{3mm}

\noindent
iii. The C-NOT gate ${\cal C}_{lm}$ is defined as
 \begin{eqnarray}
 {\cal C}_{lm} = \stackrel{12}{[+]}_l + \stackrel{12}{[-]}_l [ - \stackrel{03}{(+i)}_m 
 \stackrel{12}{(-)}_m + \stackrel{03}{(-i)}_m  \stackrel{12}{(-)}_m - 
 \stackrel{03}{(-i)}_m \stackrel{12}{(+)}_m + \stackrel{03}{(+i)}_m  \stackrel{12}{(+)}_m], 
  \label{cgate}
 \end{eqnarray}
or equivalently 
 \begin{eqnarray}
 {\cal C}_{lm} = \stackrel{12}{[+]}_l + \stackrel{12}{[-]}_l [\tau^{-}_m + \tau^{+}_m].
   \label{cgatetau}
 \end{eqnarray}

{\it Statement:} The C-NOT gate  ${\cal C}_{lm}$ if applying on  
$|\cdots 0_l \cdots 0_m \cdots \rangle$ transforms it  back to the same state,  
if applying on $|\cdots 0_l \cdots 1_m \cdots \rangle$ transforms it back to the same state. 
If ${\cal C}_{lm}$ applies on $|\cdots 1_l \cdots 0_m  \cdots \rangle$ transforms it to 
$|\cdots 1_l \cdots 1_m  \cdots \rangle$, while it transforms the state 
$|\cdots 1_l \cdots 1_m  \cdots \rangle$ to the state $|\cdots 1_l \cdots 0_m  \cdots \rangle$.

{\it Proof:} To prove this statement one only has to apply the operator ${\cal C}_{lm}$ 
on  states $|\cdots i_l \cdots i_m \cdots \rangle$ , $|\cdots i_l \cdots i_m \cdots \rangle_L$, 
$|\cdots i_l \cdots i_m \cdots \rangle_R$,  with $i_l,i_m$ equal $0$ or $1$, taking into  
 account equations from Sect. \ref{spinorstates}.

{\it Statement:} When applying $\prod^{n}_i \, {\cal H}_i$ on the $n$ qubit state 
with all the qubits in 
the state $|0_i \rangle$, we get the state $|\psi^{0}_0 \rangle $
 \begin{eqnarray}
 |\psi^{0}_0 \rangle  = \prod^{n}_i \, {\cal H}_i|0_i \rangle = 
 \frac{1}{2^{n/2}}\,\prod^{n}_i(|0_i\rangle + |1_i\rangle ).
   \label{psi0}
 \end{eqnarray}
{\it Proof:} It is straightforward to prove this statement, 
if the statement ii. of this section 
is taken into account.

\section{Useful properties of quantum gates in the technique using nilpotents and projectors}
\label{properties}

We present in this section some useful relations. % which .....

%\begin{etimize}

%\item 

\noindent
i. One easily finds, taking into account 
Eqs.(\ref{phigate},\ref{hgatetau},\ref{cgatetau},\ref{tau},\ref{taurel}), 
the relation
 \begin{eqnarray}
 {\cal R}_{\Phi_l} {\cal H}_l{\cal R}_{\vartheta_l} {\cal H}_l &=& \frac{1}{2} \{
 (\stackrel{12}{[+]}_l + e^{i \varphi_l}\stackrel{12}{[-]}_l) (1 + e^{i \vartheta_l}) + 
 (\tau^{+}_l + e^{i \varphi_l}\tau^{-}_l)(1 - e^{i \vartheta_l})\}, 
   \label{rhrh}
 \end{eqnarray}
which transforms $|i_l\rangle$ into a general superposition of $|0_l\rangle$ and 
$|1_l\rangle$ 
\begin{eqnarray}
e^{-i \vartheta_l} {\cal R}_{(\varphi_l + \pi/2) } {\cal H}_l {\cal R}_{2\vartheta_l } {\cal H}_l
|0_l\rangle &=& \cos(\vartheta_l)|0_l \rangle + e^{i \varphi_l} \sin(\vartheta_l) |1_l\rangle,\nonumber\\
e^{-i (\vartheta_l - \pi/2)} {\cal R}_{(\varphi_l -\pi/2) } {\cal H}_l {\cal R}_{2\vartheta_l } {\cal H}_l
|1_l\rangle &=& \sin(\vartheta_l)|0_l \rangle + e^{i\varphi_l} \cos(\vartheta_l) |1_l\rangle. 
\label{supl}
\end{eqnarray}

\noindent
ii. Let $|\psi_m \rangle_{p}$ be a general superposition of $p$-qubit states 
$|k \rangle_{p} = \prod_{l=1,p} 
 \,(\tau^{-}_{l})^{i_l}|0 \rangle_{l},$ with $i_l=0,1$ for a particular 
 choice of $i_l$ so that $|k \rangle_{p}$ represents any of the $2^p$ basic 
 $p$-qubit states: 
$|\psi_m \rangle_{p} =  \sum_{k=1}^{2^p} \; \alpha^{m}_k \, |k \rangle_{p}$.  
Then for $\sum_{k=1}^{2^p}\, \alpha_{k}^{m\star} \alpha^{m}_k =1$  we find that the operator 
$\hat{O}^{m}_p$ is an involution
\begin{eqnarray}
\label{pgeninv}
\hat O^{m}_{p} &=& 2|\psi^m \rangle_{p}\;\, {}_{p}\langle \psi^m| -I, \nonumber\\ 
(\hat O^{m}_{p})^{\dagger} &=& \hat O^{m}_{p}, \quad (\hat O^{m}_{p})^{2} = I. 
\end{eqnarray}
We also find that any operator $\hat P^{k}_{p} = I - 2\hat R^{k}_{p}$ with $\hat R^{k}_{p} = 
\prod_{l=1}^p \;|i_l\rangle \, \langle i_l | = 
|k \rangle_{p}\;\, {}_p\langle  k|$, with a particular choice of $i_l$, is also an involution
\begin{eqnarray}
\label{Ok0p}
\hat P^k_{ p} &=& I - 2 \prod_{l=1}^p\; \;|i_l\rangle \, \langle i_l |
= I - 2\hat R^{k}_{ p}, \nonumber\\
(\hat P^{k}_{p})^2 &=& (I - 2\hat R^{k}_{ p})^2 = I.  
\end{eqnarray}
\section{Algorithm for extracting particular states}
\label{algorithm}

Let as define a quantum  algorithm, which extracts a particular information out of a data base 
with $n$ qubits. We assume that the starting state is any superposition of  
$2^p$ states, out of which we are extracting a particular state. 
%, or a set of particular states if $p < n$. 
In the case that the starting state is a superposition of $2^p$ states with equal 
coefficients (Eq.\ref{psi0}), all of them equal    
to $2^{-p/2}$, the algorithm is known as the Grover's algorithm~\cite{grover}. 
Let $|k\rangle_p$ be a state of $p$ qubits 
\begin{eqnarray}
\label{km}
{}_{p}\langle k|m\rangle_{p} &=&\delta^{km},  
\end{eqnarray}
and let $|k_0\rangle_p$  be a particular state of p qubits, which we would like to extract out of 
an general superposition  $|\psi^m\rangle $  of 
 $2^p$ orthogonal states $|k\rangle_p$
\begin{eqnarray}
\label{psis}
|\psi^m\rangle &=& \sum_{k=1}^{2^p} \frac{\alpha^{m}_k}{\sum_{k'} \alpha^{*}_{k'} \alpha_{k'}} 
|k \rangle_{p},\nonumber\\
&=& A^{m} |k^{m}_0\perp \rangle_{p} + B^{m} |k_0 \rangle_{p},
\end{eqnarray}
with %$A^{ m \star } = A^{m}$ and $B^{ m \star } = B^{m} $ and 
\begin{eqnarray}
\label{psis1}
{}_{p}\langle k_0|k^{m}_0 \perp \rangle_{p}&=& 0,\quad I = \sum_k \,|\psi^k \rangle_{p}\;
{}_{p}\langle \psi^k | % + |k_0 \perp \rangle_{p}\;{}_{p}\langle k_0 \perp|
, \nonumber\\
|k^{m}_0 \perp \rangle_{p}&=&\sum_{k \ne k_0} \; 
\frac{\alpha^{m}_k}{\sqrt{\sum_{k' \ne k_0} \, \alpha^{m \star}_{k'} \alpha^{m}_{k'}} }\, |k\rangle_p, 
\nonumber\\ 
A^m&=& \cos \vartheta_m = \frac{\sqrt{\sum_{k' \ne k_0} \, 
\alpha_{k'}^{m \star} \alpha^{m}_{k'}}}{\sqrt{\sum_{k'} \, \alpha_{k'}^{m\star} \alpha^{m}_{k'}}}
= \sqrt{1 - \frac{\alpha_{k0}^{m \star}\, \alpha_{k0}^{m}}{\sum_{k'} \, \alpha_{k'}^{m\star} \alpha_{k'}^{m}}},\nonumber\\
B^m &=& \sin \vartheta_m  e^{i \varphi_m} 
= \frac{\alpha^{m}_{k_0}}{\sqrt{\sum_{k'} \, 
\alpha^{m\star}_{k'} \alpha^{m}_{k'}}}. 
\end{eqnarray}

Let us define the operator $\hat {\cal E}^{m}_p$
\begin{eqnarray}
\label{Eenp}
\hat {\cal E}^{m}_p &=& (2|\psi^m \rangle_p \; {}_p \langle \psi^m| -I)\,
(I-2|k_0 \rangle_p \; {}_p \langle k_0|).
\end{eqnarray}
We find $\hat {\cal E}_{p}^{m \dagger} \hat {\cal E}^{m}_p =I$. Since we can write 
\begin{eqnarray}
|\psi^m\rangle_p \; {}_p \langle \psi^m| &=& \cos^2 \vartheta_m |k^{m}_0 \perp \rangle_{p}\;
{}_{p}\langle k^{m}_0 \perp| + \sin^2 \vartheta_m |k_0  \rangle_{p}\;
{}_{p}\langle k_0 | \nonumber\\
&+& \sin\vartheta_m \cos \vartheta_m (e^{-i \varphi_m}
|k^{m}_0 \perp \rangle_{p}\; {}_{p}\langle k_0| + 
e^{i \varphi_m}
|k_0  \rangle_{p}\;{}_{p}\langle k^{m}_0 \perp|), 
\label{psim}
\end{eqnarray}
it accordingly follows
\begin{eqnarray}
\hat {\cal E}^{m}_p \,|\psi^m\rangle_p &=& [\cos 2 \vartheta_m  + 
\sin 2 \vartheta_m ( e^{i \varphi_m}
\,|k_0  \rangle_{p}\;{}_{p}\langle k^{m}_0  \perp| - e^{-i \varphi_m}
|k^{m}_0 \perp \rangle_{p}\;{}_{p}\langle k_0| )]\,|\psi^m\rangle_p \nonumber\\
&=& \cos (3 \vartheta_m )\, |k^{m}_0 \perp \rangle_{p}\; + e^{i \varphi_m}\,
\sin (3 \vartheta_m )\,|k_0 \, \rangle_{p}.
\label{gpm}
\end{eqnarray}
If we apply the operator  $\hat {\cal E}^{m}_p \;$   $j$ times, we find
\begin{eqnarray}
(\hat {\cal E}^{m}_p)^j \,|\psi^m \rangle_p &=& [\cos (2 j \vartheta_m )\;  + 
\sin (2j \vartheta_m ) \;( e^{i \varphi_m}\,|k_0  \rangle_{p}\;
{}_{p}\langle k^{m}_0  \perp| - e^{-i \varphi_m}\,|k^{m}_0 \perp \rangle_{p}\;
{}_{p}\langle k_0| )] \,|\psi^m\rangle_p \nonumber\\
&=& \cos [(2 j +1) \vartheta_m ] \; |k^{m}_0 \perp \rangle_{p}\; + e^{i \varphi_m}\,
\sin [(2j+1) \vartheta_m ]\;|k_0 \, \rangle_{p}.
\label{gpmmp}
\end{eqnarray}
It follows that $(\hat {\cal E}^{m}_p)^j$ extracts our particular state $|k_0 \rangle_{p}$
out of the initial state $|\psi^{m}\rangle_p$, if we choose $j$ 
so that
\begin{eqnarray}
\sin [(1 + 2 j) \vartheta_m  + \varepsilon]=\frac{\pi}{2}
\label{jmo}
\end{eqnarray}
for as small $|\varepsilon|$ as possible. 
If we choose the initial state  $|\psi^{m}\rangle_p$ to be just our desired state,
then $\vartheta_{m}=\pi/2$ and $j=0$. If the 
initial state has all the coefficients equal to $2^{-p/2}$, then this is the   
Grover's algorithm~\cite{grover}, provided that $
\sin \vartheta_{m} = 2^{-p/2}$.

\section{Concluding remarks}

We  presented in this paper how can the Clifford algebra technique~\cite{hn02,hn03} be used 
in quantum computers for generating quantum gates, demonstrating that the Clifford 
algebra technique\cite{hn02,hn03} makes the 
formation of quantum gates very transparent.
Although our projectors and nilpotents can as well be 
expressed in terms of the ordinary projectors and the ordinary operators 
(we chose the phases so that when going into the matrix representation the 
usually used matrices are reproduced), the elegance of the 
technique is helpful to better understand the operators appearing in the quantum gates 
and quantum algorithms. 

We present also the algorithm appropriate for extracting an information out of any  superposition 
of $2^n$ $n$ qubit states. This is a generalization of the well known 
Grover's algorithm,  allowing in principle a faster extraction of a particular information than 
the Grover's algorithm, if the initial state favours the particular state. 
Our algorithm becomes the Grover's algorithm for a very special 
choice of the initial state, out of which the information is extracting. 

The presented Clifford algebra technique (with nilpotents and projectors\cite{hn02,hn03}) 
used in this paper for quantum gates and algorithms for the  two states 
quantum bits can be quite easily generalized to   
 cases with more 
than two states qubits.

\end{document}